\documentclass[fleqn,usenatbib]{mnras}

\usepackage{newtxtext,newtxmath}

\usepackage[T1]{fontenc}

\DeclareRobustCommand{\VAN}[3]{#2}
\let\VANthebibliography\thebibliography
\def\thebibliography{\DeclareRobustCommand{\VAN}[3]{##3}\VANthebibliography}

\usepackage{graphicx}	
\usepackage{amsmath}	

\title[Constraining the nature of FRB-emitting bunches]{Constraining the nature of FRB-emitting bunches via photo-magnetic cascades}

\author[A. J. Cooper et al.]{
A. J. Cooper,$^{1,2}$\thanks{E-mail: a.j.cooper@uva.nl}
R. A. M. J. Wijers,$^{1}$
\\
$^{1}$API, Anton Pannekoek Institute for Astronomy, University of Amsterdam, Science Park 904, 1098 XH Amsterdam, the Netherlands \\
$^{2}$ASTRON, Netherlands Institute for Radio Astronomy, Oude Hoogeveensedijk 4, 7991 PD, Dwingeloo, the Netherlands
}

\date{Accepted XXX. Received YYY; in original form ZZZ}

\pubyear{2021}

\begin{document}
\label{firstpage}
\pagerange{\pageref{firstpage}--\pageref{lastpage}}
\maketitle

\begin{abstract}
We provide constraints on the nature of particle bunches that power fast radio bursts (FRBs) in the coherent curvature radiation model. It has been shown that current-induced perturbation to the motion of individual particles results in a high-energy, incoherent component of emission. We consider photo-magnetic interactions and show that the high-energy radiation can produce pairs which screen the accelerating electric field. We find that to avoid catastrophic cascades that quench emission, bunches capable of producing FRBs must have a modest density $n_e \approx 10^{13-14}\, {\rm cm^{-3}}$, and likely propagate along field lines with large curvature radii, $\rho > 10^8 \, {\rm cm}$. This rules out rapidly rotating magnetars as FRB sources within the coherent curvature radiation model.
\end{abstract}



\section{Introduction}
The coherent emission of particles is required for short, luminous radio transients with brightness temperatures too high to be explained by the sum of incoherent particle radiation. Highly magnetized neutron stars are known to produce extremely bright coherent radio emission \citep{Hankins2007,Bochenek2020,CHIME2020}. Fast radio bursts (FRBs) are extremely bright millisecond duration radio transients \citep{Lorimer2007,Thornton2013,Spitler2014,Petroff2016,Petroff2021,CHIME2021}, and extragalactic highly magnetized neutron stars known as magnetars could be the source of this enigmatic emission \citep{Keane2012,Popov2013,Katz2016,Beloborodov2017,Michilli2018,Margalit2018,Wadiasingh2019}, possibly with properties different to those of known galactic magnetars \citep{Beniamini2020}.
\par
Although there are many theories of the origin of FRBs, two leading theories have emerged; synchrotron maser emission from relativistic shocks \citep{Lyubarsky2014,Ghisellini2017,Waxman2017,Metzger2019,PLotnikov2019}, and coherent curvature radiation from the magnetosphere \citep{Cordes+2016,Kumar2017,Katz2018,Ghisellini2018,Lu2018,Yang2018,KumarBosnjak2020} which was originally formulated to explain pulsar emission \citep{Sturrok1971, RudermanSutherland1975} and is the focus of this paper.
\par
In the coherent curvature radiation model of FRBs, particles are continuously accelerated by an electric field with a component parallel to the neutron star magnetic field lines. It is often assumed that the unscreened, transient electric field is triggered by a stochastic event such as magnetic field reconnection or crustal fracturing of the neutron star surface. Accelerated particles propagate along magnetic field lines, as motion perpendicular to the field lines is rapidly radiated due to the strong magnetic field. As the particles propagate they emit curvature radiation due to particle acceleration transverse to their velocity as they stream along curved magnetic field lines. The spatial proximity of particles and their relativistic velocities means emission can be coherent.
\par
Coherent curvature radiation can explain many observed properties of both pulsar emission and FRBs (e.g. \citealt{Gil2004,Kumar2017,Yang2018}), but has been the focus of much criticism (e.g. Section 5 in \citealt{Melrose2021}). The primary difficulty facing the model is focused on the formation and continuation of charged bunches, as inhomogeneity is required for coherent emission. Early proponents of the theory suggest that bunching of particles such that emission is coherence could arise naturally from the two-stream instability or due to the radiation-reaction force \citep{GoldreichKeely1971,ChengRuderman1977,BenfordBuschauer1978}, but it is noted in \cite{GoldreichKeely1971} that this is unstable to differential particle velocities in a bunch. Furthermore, the bunch growth timescale for typical particle distributions may be too long to produce short radio bursts and observed pulsar emission \citep{Melrose1992}. Assuming bunches are formed, understanding how they persist despite radiation and electrostatic repulsion presents more challenges \citep{Lyutikov2021}. At the least, electrostatic repulsion implies large lorentz factors \citep{Katz2020}, suggesting FRBs observed at GHz frequencies may propagate on open field lines (see Eq. \ref{eq:gammanu_c_curv}). Further study is required to understand the mechanism and stability of bunch formation and persistence, but this paper focuses on the consequences of bunch radiation. 

\par
It has been shown in \cite{Cooper2021} that coherent curvature radiation bunches also emit high-energy radiation due to the magnetic field component associated with the current of coherently radiating particles. The high-energy emission depends on the source parameters and could explain X-ray emission coincident with transient coherent radio emission from the Crab Pulsar \citep{Enoto2021} and SGR 1935+2154 \citep{Mereghetti2020,Li2021,Ridnaia2021}. At extra-Galactic distances from which FRBs radiate, the high-energy emission may not be observable but could significantly affect the source and the acceleration of particles. In this work, we look at the implications of high-energy emission in magnetized environments where $\epsilon B \gtrsim 2 m_e c^2 B_c$ where $\epsilon$ is the high-energy photon energy, $B$ is the local magnetic field, and $B_c \approx 4.4 \times 10^{13} \,$G is the quantum critical magnetic field. We focus on coherent emission powering FRBs which are expected to originate in magnetic fields $B > B_c$ \citep{Kumar2017}, although our results are applicable to coherent emission in $B < B_c$ regime (see Section. \ref{sect:discussion}).
\par
Determining the maximum size of coherent bunches is usually understood theoretically by arguments of causality, or due to the coherence constraints. In \cite{KumarBosnjak2020}, the authors identify three limitations to the size of bunches transverse to the bunch velocity, $R_t$. First, causality dictates that $R_t \leq ct$ where t is the time from the trigger from which acceleration begins. Second, coherence requires velocity and acceleration vectors of all particles to be confined to an angle $\gamma^{-1}$, which implies $R_t \leq \frac{R}{\gamma} = 4 \times 10^{3} \; R_6 \, \rho_{8}^{-1/3} \, \nu_{\rm obs, 9}^{-1/3} \: {\rm cm}$.  \footnote{We use convenient notation $X_{\rm n} \equiv X/10^n$ throughout.} Here $R$ is the propagation distance of the bunch from the trigger point, $\rho$ is the curvature radius of the magnetic field line and we have made use of Eq. \ref{eq:gammanu_c_curv}. In \cite{Lu2018}, the authors suggest $R$ should correspond to the radiation formation length scale $\rho/\gamma$ such that the maximum transverse bunch size is $\frac{\rho}{\gamma^2} \approx 2 \times 10^{3} \; \rho_{8}^{1/3} \, \nu_{\rm obs,  9}^{-2/3} \: {\rm cm}$. The last constraint discussed in \citep{KumarBosnjak2020} is that particles should be confined to the transverse width of the propagating Alfv\'en wave. This is specific to their model of bunch formation, and is assumed by the authors to be less constraining than the first and second conditions. In this paper we introduce a new stringent constraint to the transverse size of coherent bunches due to pair cascades initiated by high-energy radiation emitted by coherent bunches. Using this constraint, we explore the nature of bunches that can emit FRB luminosity radio emission. 

\section{Pair cascades initiated by photo-magnetic interaction}
\label{sect:2}
The acceleration and coherent radiation of charged bunches requires continuous acceleration to offset significant radiative losses. We consider the possibility that high-energy radiation emitted by particles in coherent bunches leads to cascades of pair production, which subsequently screen the accelerating electric field $E_{\parallel}$ on short timescales. 
\par
High-energy photons can interact with the magnetic fields to either split into two photons, or to produce a pair of charged particles \citep{Erber66,Adler1970,DaughertyHarding83}. One-photon pair production is a crucial part of polar cap model of pulsar emission, in which gamma-ray curvature photons interact with the field to produce secondary pairs \citep{Sturrok1971,RudermanSutherland1975} which subsequently emit coherent radio emission. The photon splitting process has been invoked to explain a lack of radio emission from high-field pulsars \citep{Baring1998}. Following the literature, we introduce a parameter $\chi$ for convenience such that:
\begin{equation}
    \chi = \frac{\epsilon }{2 m_{\rm e} c^2} \frac{B}{B_c} \sin(\theta_{\rm \, kB})
\end{equation}
Where $\epsilon = h \nu$ is the photon energy and $\theta_{\rm \, kB}$ is the angle between the photon velocity vector and the magnetic field. Following \cite{DaughertyHarding83}, we can write down an approximation for the photon attenuation factor due to one-photon pair production \footnote{We note here that we do not include the factor $f(\epsilon, B)$ that is used in the more accurate analytic expression as described in \cite{DaughertyHarding83} for $T_{\rm pp}$, as there is no significant difference for our cases of interest.}:
\begin{equation}
    T_{\rm pp} = \begin{cases} 0.23 \alpha \bigg(\dfrac{m_e c}{\hbar}\bigg) \bigg(\dfrac{B}{B_c}\bigg) \sin(\theta_{\rm \, kB}) \exp\big(\frac{-4}{3 \chi}\big) \; {\rm cm^{-1}} & \rm  \: \: \epsilon > 2 m_e c^2 \\
0 & \rm  \: \:  \epsilon < 2 m_e c^2
    \label{eq:pairproduction}
    \end{cases}
\end{equation}
where $\alpha$ is the fine structure constant. For attenuation due to photon splitting, we use the following approximation \citep{Adler1970,Baring1997}:
\begin{equation}
T_{\rm sp} = \bigg( \dfrac{\alpha^3}{10 \pi^2}\bigg) \bigg( \dfrac{19}{315}\bigg)^{2} \bigg(\dfrac{m_e c}{\hbar}\bigg) \bigg(\dfrac{B}{B_c}\bigg)^{6} \ell(B_c) \bigg(\frac{\epsilon}{m_e c^2} \bigg)^{5} \sin^{6}(\theta_{\rm \, kB}) \; {\rm cm^{-1}}
\label{eq:splitting}
\end{equation}
\begin{figure}
    \centering
    \includegraphics[width=0.48\textwidth]{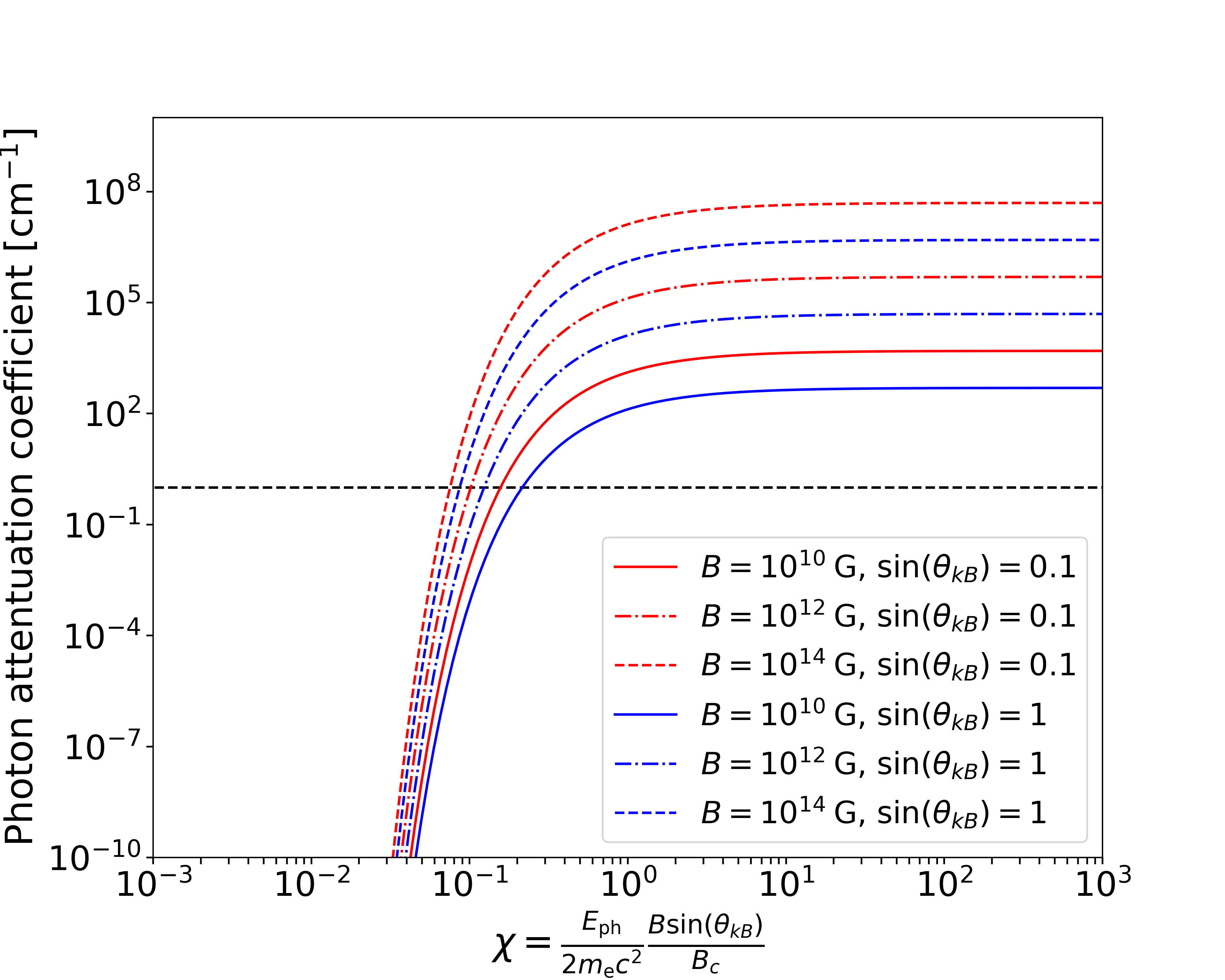}
    \caption{Photon attenuation coefficient due to one-photon pair production as a function of $\chi$ for various magnetic field strengths and photon-magnetic field angles $\theta_{\rm \, kB}$. If $\chi > 1$, photons always interact with the field to produce pairs for the parameters we show.}
    \label{fig:photon_attenuation_chi}
\end{figure}

Here $\ell(B_c) = 1$ if $B < B_c$, and $\ell(B_c) \propto B^{-6}$ if $B > B_c$, cancelling dependence on $B$ in the strongest field limit.
\begin{figure}
    \centering
    \includegraphics[width=0.48\textwidth]{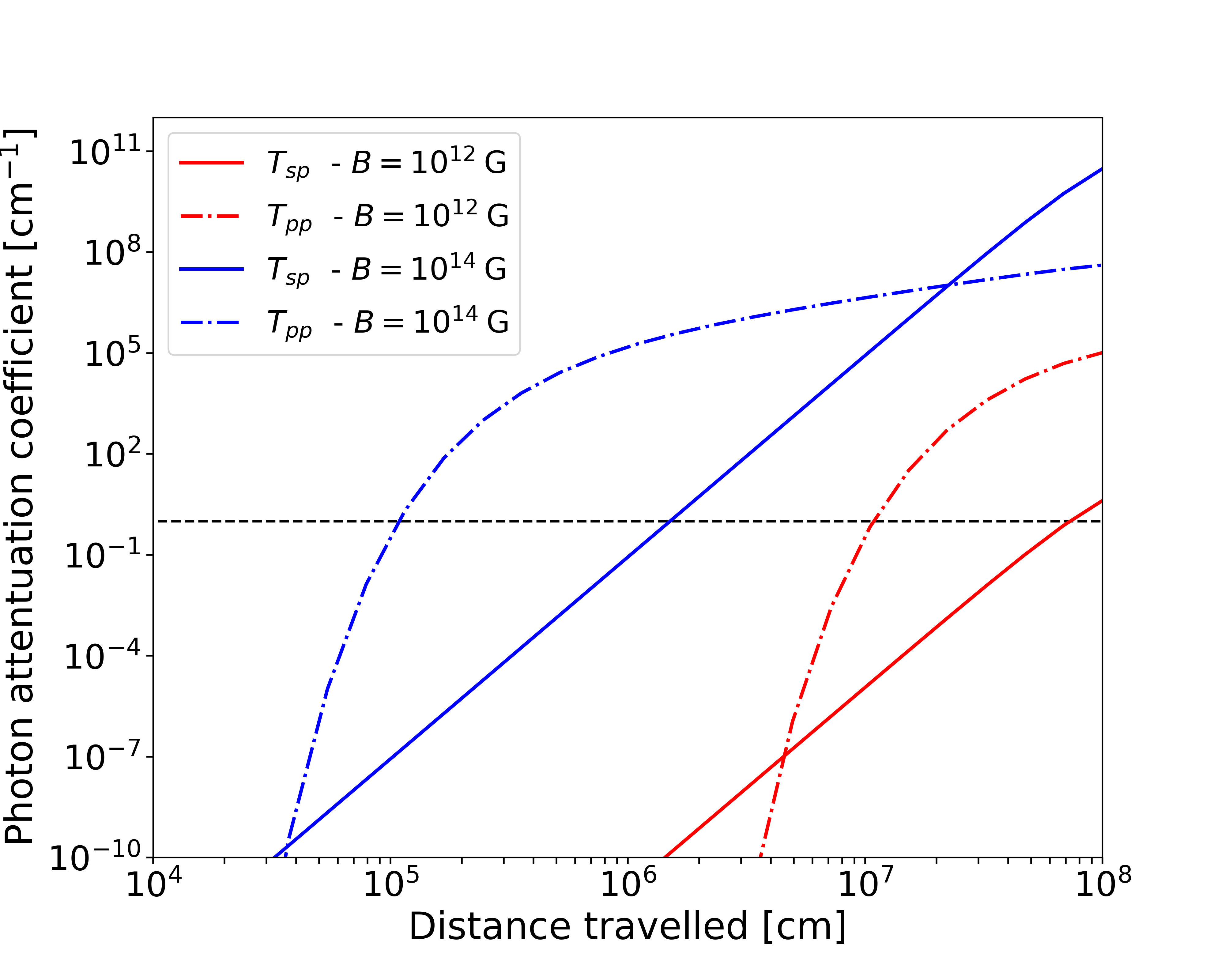}
    \caption{Here we show the photon attenuation due to photon splitting and one-photon pair production channels as a function of distance traversed for two magnetic field strengths. We have used $\rho = 10^8 \,$cm and $\epsilon = 10 \, m_e \, c^2 \approx 5 \,$ MeV. In each case, efficient attenuation due to one-photon pair production occurs earlier in the photon's path assuming that $\sin{\theta_{\rm \, kB}}$ increases proportionally to the distance the photon has propagated.}
    \label{fig:photon_attenuation}
\end{figure}
\begin{figure}
    \centering
    \includegraphics[width=0.48\textwidth]{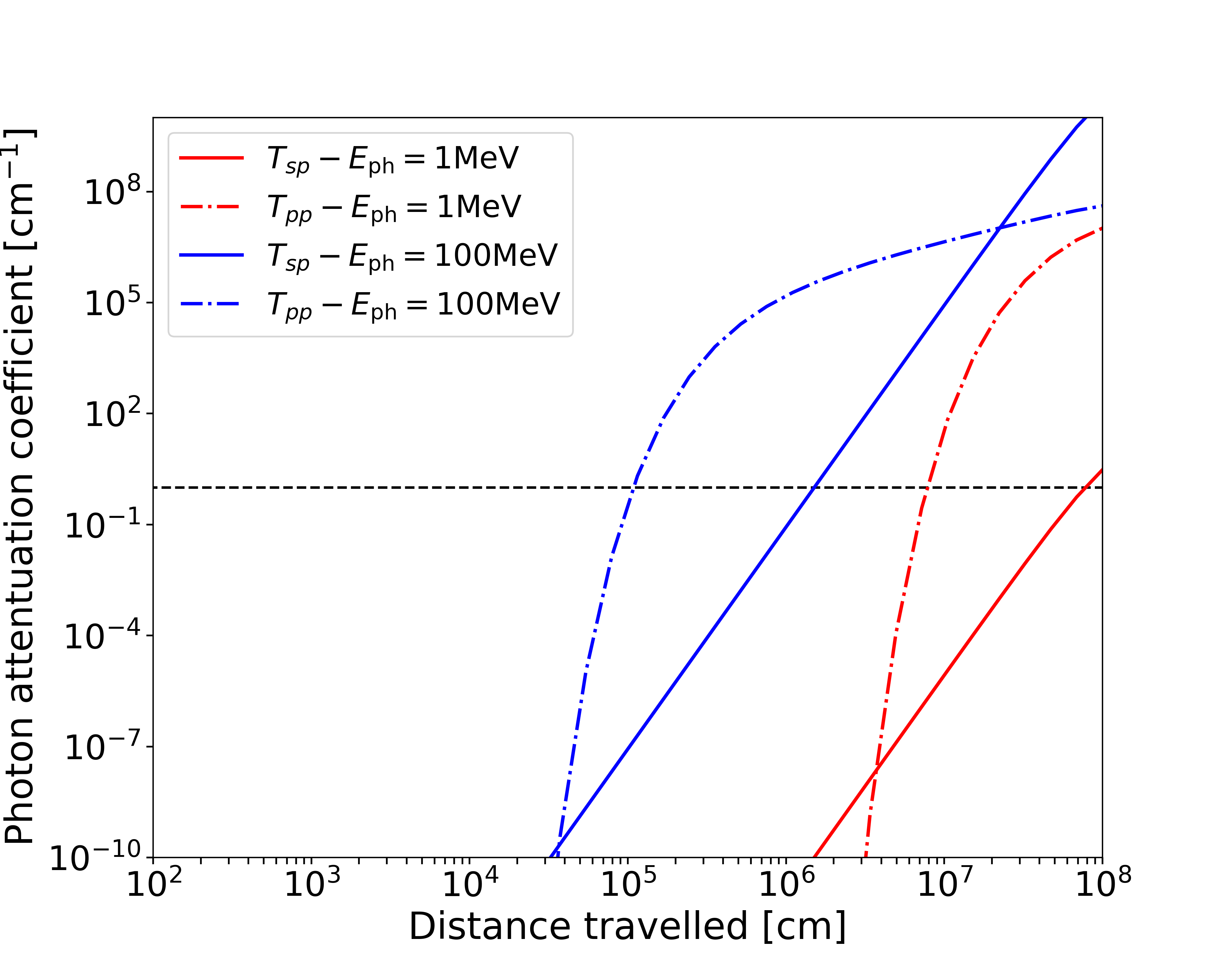}
    \caption{Here we show the photon attenuation as a function of distance traversed due to photon splitting and one-photon pair production channels for different photon energies. We have used $\rho = 10^8 \,$cm and $B = 10^{14} \,$G.}
    \label{fig:photon_attenuation_2}
\end{figure}
If the condition $\chi > 1$ is met, photons interact with the magnetic field, as seen in Fig. \ref{fig:photon_attenuation_chi} where we show the one-photon pair production channel. We can see from Figs. \ref{fig:photon_attenuation} \& \ref{fig:photon_attenuation_2} that in general one-photon pair production is the dominant process if $\epsilon > 2 m_e c^2$ and assume that the angle to the field $\sin(\theta_{\rm \, kB})$ increases linearly as $d/\rho$ where $d$ is the distance traversed by the photon. This is expected as the photon is emitted tangentially to the magnetic field and follows a straight line. We note that photon splitting can be dominant for photons just below the pair production threshold, but as these photons cannot produce pairs due to energy constraints this is unimportant for pair-photon cascades.
\par
Pairs produced via this channel are preferentially produced in the ground lowest energy state, and possibly form a very short-lived bound state known as postrionium \citep{Usov1995}. Assuming the electric field persists, the new pairs are rapidly accelerated to high-energies, until they reach a terminal lorentz factor where radiative losses are balanced. In this time, they
emit radiation, which in turn can interact with the magnetic field. If these processes occur on short enough spatial and temporal timescales such that it occurs a few times within the local electric field a photon-pair cascade occurs. If such a cascade occurs on short enough timescales, the electric field is effectively screened due to the large current associated with the back-reaction due to the acceleration of charged particles and without continuous acceleration the coherent radiation of particles will cease. 
\subsection{High-energy emission from coherent bunches}
Let us assume a bunch of $N$ charged particles are continuously accelerated along a magnetic field line $B_{\parallel}$ by an electric field $E_{\parallel}$, with a number density $n_e$, transverse extent $R_t$ and longitudinal extent $R_l$. The current associated with the accelerated bunch is approximately $J = 2 n_{\rm e} q c$, inducing a magnetic field $B_{\perp} = \dfrac{4 \pi R_t J}{c} = 8 \pi R_{t} q n_{\rm e}$. The particles follow the new total field line $B = B_{\parallel} + B_{\perp}$, and therefore radiate both coherent curvature emission and high-frequency incoherent emission. As in \cite{Cooper2021}, the critical frequency and emitted (per particle) power of synchrotron-like high-energy radiation produced by coherent bunches is (where $\alpha \ll 1$):
\begin{equation}
    \begin{split}
        \nu_{\rm c,sync} &= \frac{3}{2} \gamma^3 \omega_B \sin(\alpha) \\
        &= \frac{3}{2} \gamma^3 \bigg(\frac{q B_{\parallel}}{\gamma m_{\rm e} c}\bigg) \frac{8 \pi q n_{\rm e} R_t}{B_{\parallel}} \\
        &= \frac{12 \pi q^2  n_{\rm e} \gamma^2 R_t}{m_{\rm e} c} \\
    \end{split}
    \label{eq:nu_c_sync}
\end{equation}
\begin{equation}
    \begin{split}
    P_{\rm sync} &= \frac{1}{4} \pi c \sigma_T B_{\parallel}^2 \gamma^2 \sin(\alpha)^2 \\
    &= 16 \pi^3 c q^2 \sigma_T R_t^2 n_e^2 \gamma^2
    \end{split}
    \label{eq:power_sync}
\end{equation}
Assuming that coherent curvature radiation is also observed from the bunch, we can rewrite $\gamma$ in terms of the observed wavelength of coherent emission. For coherent curvature radiation the critical frequency is defined as: $\nu_{\rm c,curv} = \frac{3 c \gamma^3}{4 \pi \rho}$. Therefore in terms of the curvature radius $\rho$ and the observing frequency of the coherent radio emission $\nu_{\rm obs, curv}$, $\gamma$ is:
\begin{equation}
    \begin{split}
        \gamma &= \bigg(\frac{4 \pi \rho \nu_{\rm obs, curv}}{3c} \bigg)^{1/3} \\
        &= 240 \; \rho_8^{1/3} \, \nu_{\rm obs, curv,9}^{1/3} 
    \end{split}
    \label{eq:gammanu_c_curv}
\end{equation}
It is difficult to estimate the number density, $n_e$, of bunches emitting coherent curvature radiation, and estimates of $n_e$ based on observed FRBs often are degenerate with other model parameters such as the curvature radius of field lines $\rho$. One method to theoretically estimate the bunch number density is to parametrize in terms of Goldreich-Julian density \citep{Goldreich1969,Kumar2017} such that:
\begin{equation}
\begin{split}
    n_e &= \xi n_{\rm GJ} = \frac{2 \xi B_{\parallel}}{q c P} \\
    &= 10^{15}\; \xi_1 \, B_{\parallel, 15}\, P_0^{-1}
\end{split}
    \label{eq:goldreich}
\end{equation}
Where the pair multiplicity factor $\xi$ tells us how much larger the number density of the bunch is compared to the Goldreich-Julian density due to the local magnetic field. An overdensity of electrons above the Goldreich-Julian density is generally required for radio emission originating in the magnetosphere \citep{YangZhang2018}, and this might be larger (e.g. $\xi \gg 1$) for extreme transient emission like FRBs. In many FRB models the value is somewhat unconstrained and often depends on the trigger mechanism and on the efficiency of pair creation. In \cite{Beloborodov2020}, a value of $\xi = 10^3$ is suggested due to pair creation sustained by the magnetar's twisted field lines.
\par
We can solve Eq. \ref{eq:nu_c_sync} for the value of $R_t$ for the threshold case where synchrotron photons are emitted with an energy such that pair production is allowed, $\epsilon > 2 m_e c^2$:

\begin{equation}
    \begin{split}
        h \nu_{\rm c,sync} &> 2 m_e c^2 \\
        R_t &> 13 \; \big( n_{e,15}^{-1} \, \rho_8^{-2/3} \, \nu_{\rm obs, curv, 9}^{-2/3} \big) \: {\rm cm}
    \end{split}
    \label{eq:photons_above_threshold}
\end{equation}
The above equation tells us that for the given parameter set, bunches with a transverse size larger than $13 \,$cm will produce high-energy radiation capable of producing additional pairs via photon-magnetic field interaction. It is argued in Section \ref{sect:opacity} \& \ref{sect:cascades} that in some cases, this size constitute the maximum transverse bunch size due to photon-pair cascades that screen the field. In Section \ref{sect:consequences} we compare this to the two other aforementioned constraints on $R_t$ due to causality and from \cite{KumarBosnjak2020}. 

\par
\begin{figure}
    \centering
    \includegraphics[width=0.48\textwidth]{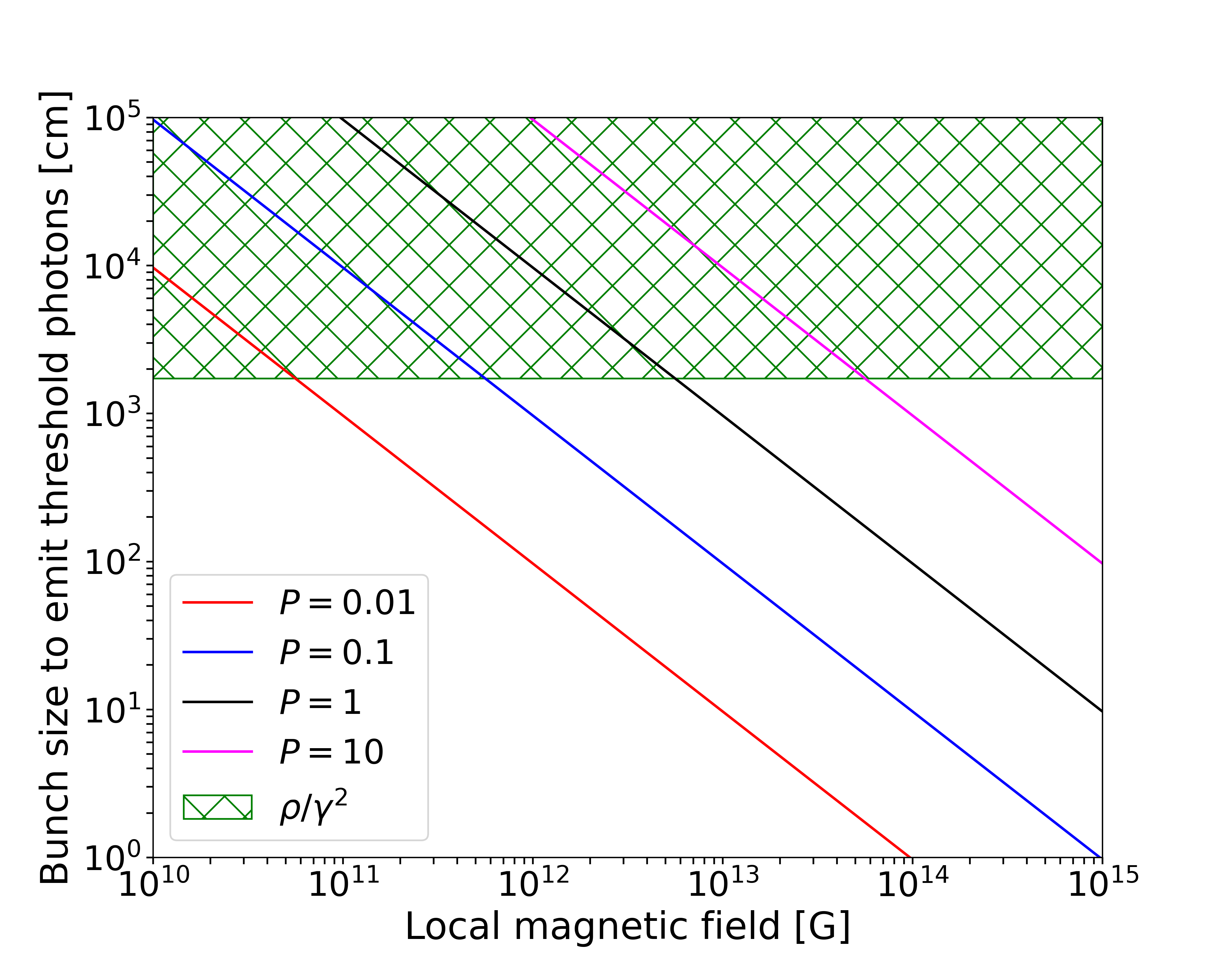}
    \caption{We plot Eq. \ref{eq:photons_above_threshold} for various values of the local magnetic field and pulsar period. The lines refer to the bunch size that corresponds to the minimum required size to emit high-energy radiation above the pair production threshold. Other parameters have fidicual values: $\xi = 10$,  $\rho = 10^8 \,$cm \& $\nu_{\rm obs, curv} = 10^9 \,$Hz. In green we show the constraint  $R_t < \dfrac{R}{\gamma} = \dfrac{\rho}{\gamma^2}$ \citep{KumarBosnjak2020,Lu2018}.)}
    \label{fig:bunchsize}
\end{figure}

One-photon pair production occurs efficiently as long as $\epsilon > 2 m_e c^2$ and $\chi > 1$, and therefore we can also derive constraints for neutron stars with magnetic fields less than $B_c$. Using Eq. \ref{eq:photons_above_threshold}, we show the size of bunches that can produce photons with $\epsilon = m_e c^2$ for various magnetic fields and periods in Fig. \ref{fig:bunchsize}. 
\subsection{One-photon pair production screening the electric field}
\label{sect:opacity}
For bunches to initiate cascades we not only require that they produce photons capable of pair production, but also that the photons interact with the magnetic field while still in the vicinity of the electric field that accelerates the bunches. There are two specific constraints.
\par
In the first, the distance $d$ the photon propagates before interaction with the magnetic field must be smaller than $l_{\rm l}$, the longitudinal (i.e. direction of particle acceleration) length scale of the acceleration region. In the coherent curvature radiation model of FRBs, the accelerating field is often assumed to be triggered by a crustal fracture event on the surface of the neutron star, or a reconnection event close to the surface. Given this, we could expect that $l_{\rm l} < R_{\rm NS} \approx 10^6 \; {\rm cm}$. In \cite{Kumar2017}, the longitudinal size of the accelerating current sheet is suggested to be $l_{\rm l} \gtrsim \frac{\rho}{\gamma} \approx 4 \times 10^{5} \rho_8^{4/3} \, \nu_{\rm obs,9}^{1/3}$. We assume $l_{\rm l} = 10^{6} \,$cm throughout the rest of this work for simplicity. 
\par
In the second constraint, we require that the magnetic field lines, from which high-energy photons are tangentially emitted, have not diverged so much such that the photo-magnetic interaction site is on a different set of field lines which do not accelerate bunches. After propagating a distance $d$, the transverse distance between the photon and magnetic field line from which it was tangentially emitted is approximately $l_{\rm t} = d \sin(d/\rho) \approx \frac{d^2}{\rho}$. For the field accelerating the bunches to be shorted out, this distance should be smaller than the transverse bunch size $R_t$ such that the secondary particles are accelerated within bunch acceleration region. Therefore we require $d < (R_t \rho)^{1/2}$. 
\par
Given these two conditions, we require that to produce cascades, one-photon pair production before traversing a length scale $l$ where:
\begin{equation}
    l = {\rm min( l_{\rm l}, l_{\rm t})} = {\rm min ((R_t \rho)^{1/2}, 10^6 {\rm cm})}
    \label{eq:l_definition}
\end{equation}
\par
The opacity due to one-photon pair production is given by:
\begin{equation}
    \sigma_{\rm pp} =  \int_{0}^{l} T_{\rm pp} \,dl
    \label{eq:opacityintegral}
\end{equation}
Therefore if $T_{\rm pp} \gtrsim l^{-1}$, pair production occurs within the field which accelerates the bunches, assuming $T_{\rm pp}$ does not significantly evolve over the photon path. Photons propagate away from the neutron star, and therefore the local magnetic field will decrease along the photons path. However, Eq. \ref{eq:l_definition} states that the photon path is not larger than the neutron star length scale $r = 10^6 \,$cm, so the assumption that $T_{\rm pp}$ does not significantly evolve is approximately true. Cascades only occur within $l$, therefore we calculate the value of $\sin(\theta_{\rm \, kB})$ as follows:
\begin{equation}
    \sin(\theta_{\rm \, kB, max}) = \begin{cases} 1 \; & \rm  \: \: \rho \leq l \\
\frac{l}{\rho} & \rm  \: \:  \rho > l
    \label{eq:double_condition}
    \end{cases}
\end{equation}
Where we have assumed that the photon is emitted tangent to the magnetic field as the angle to the field grows as $\theta_{\rm \,kB} \approx \frac{d}{\rho}$, where $d$ is the distance the photon has propagated. 
\subsection{Cascade timescales}
\label{sect:cascades}

The duration of which one bunch of particles is visible in the observer frame is approximately $\nu_{\rm obs}^{-1}$ due to its longitudinal size, which is much smaller than the observed FRB duration. Therefore up to $10^6$ consecutively visible bunches are required to explain millisecond FRB emission \citep{Lu2018}. The accelerating electric field must therefore be sustained for the duration of the FRB, possibly by a reconnection event that lasts for the FRB duration \citep{Kumar2017}. By considering the relevant processes, we can compute the timescale upon which photon-pair cascades occur in the observer frame, which are initiated by incoherent emission from bunches. Firstly, the timescale to accelerate particles to the lorentz factor expected for coherent curvature radiation can be expressed as:
\begin{equation}
   \begin{split}
        \tau_{\rm acc} &= \frac{\gamma m_e c^2}{e E_{\parallel} c} = \frac{m_e c}{q E_{\parallel}} \bigg(\frac{4 \pi \rho \nu_{\rm obs, curv}}{3 c}\bigg)^{1/3} \\
        &= 10^{-13} \; \big(E_{\parallel, \rm 8}^{-1} \, \rho_8^{1/3} \, \nu_{\rm obs, curv, 9}^{1/3} \big) \: {\rm s }
    \end{split}
    \label{eq:tacc}
\end{equation}
Where we have used Eqs. \ref{eq:gammanu_c_curv} \& \ref{eq:goldreich}, and have assumed an electric field strength required for coherent curvature bunches to radiate FRB strength bursts (e.g. \citealt{Kumar2017,Wang2019}). Next we can look at the timescale it takes for a lepton in the coherent bunch to produce high-energy photons using Eqs. \ref{eq:nu_c_sync} \& \ref{eq:power_sync}:
\begin{equation}
    \begin{split}
    \tau_{\rm ph} &= \frac{h \nu_{\rm c, sync}}{P_{\rm sync}} = \frac{6 \pi \gamma^2 h q B_{\parallel} \sin(\alpha)}{m_e c^2 \sigma_T \gamma^2 B_{\parallel}^2 \sin^2(\alpha)}\\
    &= \frac{3 h}{4 \pi^2 m_e c^2 \sigma_T n_{\rm e} R_t} = 6 \times 10^{-13} \; \big(\rho_8^{2/3} \, \nu_{\rm obs, curv, 9}^{2/3} \big) \: {\rm s}  \\   
    \end{split}
    \label{eq:tph}
\end{equation}
This is the approximate timescale for one high-energy photon to be emitted at the critical synchrotron frequency by an accelerated electron or positron. Lastly, we can look at the timescale upon which photons produce pairs: the timescale for which a photon must propagate to subtend an angle to the magnetic field $\theta_{\rm \, kB}$ large enough such that $\chi > 1$. As mentioned in Sect. \ref{sect:opacity} we require that photons interact with the field within a length scale $l < 10^6$cm. Therefore if successful cascades occur, they do so on a maximum timescale:
\begin{equation}
    \tau_{\rm pp} = \frac{l}{c} = 3 \times 10^{-5} \; \big(l_{6} \big) \: {\rm s}  \\   
    \label{eq:tpp}
\end{equation}
This timescale is shorter than typical FRB durations, and therefore we don't expect a case in which cascades are initiated but on a long enough timescale such that an FRB is visible before the field is screened. We discuss shorter sub-millisecond bursts in Sect. \ref{sect:discussion}. In Section \ref{sect:consequences}, we assume cascades occur and the field is screened if $\sigma_{\rm pp} > 1$ within the required length scale $l$.

\subsection{Cascade model}
Each high-energy photon must propagate for a time $\tau_{\rm pp}$ before the angle to the magnetic field $\theta_{\rm \, kB}$ is large enough such that it decays to a pair which we refer to as secondary pairs. During this time, the primary charged particle will continue emitting high-energy photons, such that each primary lepton produces approximately $\frac{\tau_{\rm pp}}{\tau_{\rm ph}}$ photons before the first photon decays to pairs. To simplify, we can model the cascade as occurring in a number of distinct steps, in which after each round of pair production and acceleration which takes a time $\tau = \tau_{\rm acc} + \tau_{\rm ph} + \tau_{\rm pp} \approx \tau_{\rm pp}$, the number of photons will increase by a factor of $\frac{2 \tau_{\rm pp}}{\tau_{\rm ph}}$.
\par
Our calculations  assume that the accelerated secondary pairs will also participate in coherent emission and therefore radiate high-energy emission as defined by Eqs. \ref{eq:nu_c_sync} \& \ref{eq:tph}. If this is not the case, the pairs are limited only by their incoherent curvature losses, and are accelerated to $\gamma = 2 \times 10^8 \; E_{\parallel, 8}^{1/4} \, \rho_8^{1/2}$. In this case, the pairs still radiate photons above the pair production threshold and therefore the cascade still proceeds. In fact we will see that the field is screened before the photons produced by secondary pairs have time to interact with the field. We note that the cascade only begins after a time $\tau_{\rm pp}$ from the initial trigger, as the first photons must propagate for $d = \tau_{\rm pp} c$ before one-photon pair production occurs. In this approximation, the number density of pairs accelerated increases as a function of time as:
\begin{equation}
    n_e(t) = n_e(0) \bigg(\frac{2 \tau_{\rm pp}}{\tau_{\rm ph}}\bigg)^{\frac{t}{\tau_{\rm pp}}} \: {\rm cm^{-3}}
    \label{eq:n_e(t)}
\end{equation}
Where $n_e(0)$ denotes the initial number density of pairs. Given that $\dot{E}_{\parallel} = -4 \pi J = -8 \pi n_e(t) c q$, we can calculate the change of the field $E_{\parallel}$ due to a number of secondary pairs accelerated:

\begin{equation}
\dot{E}_{\parallel} = - 8 \pi c q n_e(0) \bigg(\frac{2 \tau_{\rm pp}}{\tau_{\rm ph}}\bigg)^{\frac{t}{\tau_{\rm pp}}} \\
\end{equation}
We can approximate the screening time as the time for one e-folding reduction in the electric field is given by $\tau_{\rm screen} \approx \frac{E_{\parallel}}{\dot{E}_{\parallel}}$ such that the field is screened when the following inequality is satisfied:
\begin{equation}
    \begin{split}
        t \dot{E}_{\parallel} &> E_{\parallel}\\
        {t \bigg(\frac{2 \tau_{\rm pp}}{\tau_{\rm ph}}\bigg)^{\frac{t}{\tau_{\rm pp}}}} &> \frac{E_{\parallel}}{8 \pi c q n_e(0)}
    \end{split}
\end{equation}
The solution to the above equation is:
\begin{equation}
   t > \frac{\tau_{\rm pp} W\bigg(\frac{E_{\parallel} \log(\frac{2 \tau_{\rm pp}}{\tau_{\rm ph}})}{4 \pi q c n_e(0) \tau_{\rm pp}}\bigg)}{\log(\frac{2 \tau_{\rm pp}}{\tau_{\rm ph}})}
\end{equation}
Where $\log$ is the natural logarithm, and $W(x)$ is the product logarithm function. Using the fiducial values we have assumed, the time to screen the field is found to be $t \approx 10^{-3} \tau_{\rm pp}$. This is due to the exponential nature of the cascade and the accelerating field is always screened before a time $\tau_{\rm pp}$ for a wide range of reasonable parameters. This means that the acceleration of a portion of secondary pairs produced by the first interacting photons provide enough current to screen the electric field. We note that although in this simplification the cascade occurs on timescales $\ll \tau_{\rm pp}$, the minimum time for the cascade to begin is approximately $\tau_{\rm pp}$ as the photons do not interact with the magnetic field before they have propagated for $\tau_{\rm pp}$. We conclude that cascades always quench emission on sub-millisecond timescales bunches produce photons with energies such that $T_{\rm pp} > l^{-1}$ where $l$ is defined in Eq. \ref{eq:l_definition}.
\subsubsection{Radiation by secondary particles}
Before the field is fully screened, a portion of the secondary pairs are also accelerated and emit radiation. The total number of accelerated secondaries is calculated by inserting the time to screen the field ($t \approx 10^{-3} \tau_{\rm pp}$) into Eq. \ref{eq:n_e(t)}. We find the number of secondary pairs accelerated is just a few percent of the initial number density $n_e(0)$ for fiducial parameters, and therefore do not expect a substantial luminosity increase due to the acceleration and radiation of secondary pairs.

\section{Consequences for FRB luminosity and spectral width}
\label{sect:consequences}
\begin{figure}
    \centering
    \includegraphics[width=0.48\textwidth]{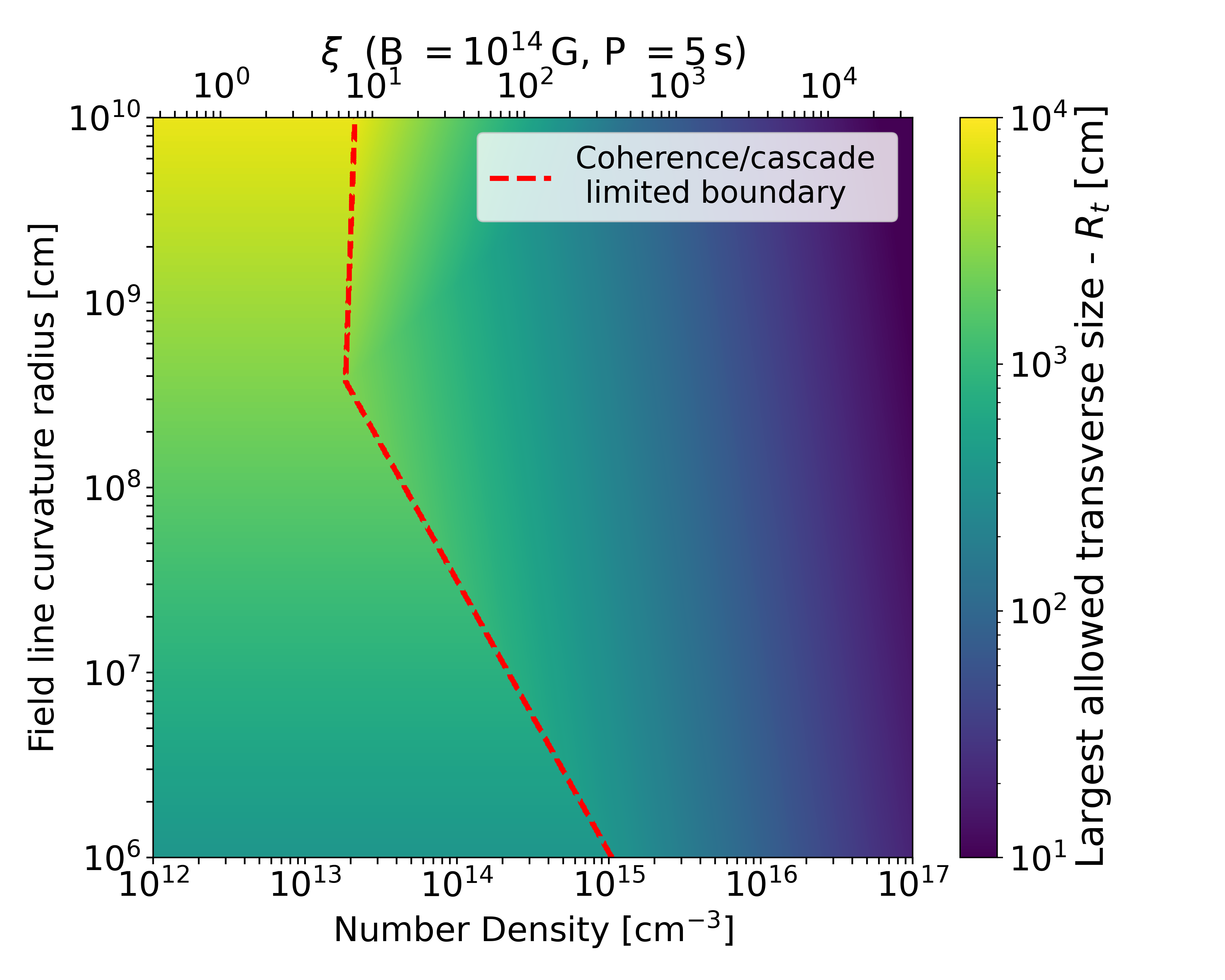}
    \caption{The maximum transverse size of bunches emitting coherent curvature radiation as a function of bunch density $n_e$ and field line curvature radius $\rho$. We include the number density in terms of the multiplicity parameter $\xi$ along the top x-axis for typical magnetar parameters for reference. We have assumed $B = 10^{14} \, {\rm G}$, $\nu_{\rm obs, curv} = 10^9 \, {\rm Hz}$ and $l = {\rm min} ((R_t \rho)^{1/2},10^6 {\rm cm})$. To the right of red line, the cascade condition outlined in this paper (Eqs. \ref{eq:opacityintegral} \& \ref{eq:double_condition}) is more constraining for the bunch size. To the left of the line the requirement that $R_t < \frac{\rho}{\gamma^2}$ is more constraining.}
    \label{fig:2d_bunch_size}
\end{figure}

We have shown that high-energy photons emitted from bunches of coherently radiating particles can interact with the magnetic field and produce pair cascades which screen the field and prevent emission. Therefore if a transient powered by coherent curvature radiation emits for longer than a few microseconds ($t > \frac{l}{c}$), the transverse extent of radiating bunches must be small enough such that either: photons above the pair production threshold are not emitted; or such that they do not interact with the field within the vicinity of the accelerating electric field. In Fig. \ref{fig:2d_bunch_size}, we show the maximum transverse bunch size as a function of the bunch number density, $n_e$, and the curvature radius of field lines, $\rho$. We calculate this value by finding the largest allowed transverse size $R_t$ such that cascades do not quench emission, which occurs if $T_{\rm pp} > l^{-1}$ (see Section \ref{sect:2}). For smaller values of $n_e$ to the left of the red dashed line the coherence constraint that the maximum bunch size should be less than $\rho/\gamma^2$ yields a smaller value. For number densities above approximately $10^{14} \,$cm, the maximum transverse size derived in this work is more constraining. In the following subsections we discuss the implications of these limitations in terms of observable properties of FRBs.
\subsection{Spectral width of coherent emission}
\label{sect:spectral}
The limited spectral width of FRBs is one of their most outstanding properties (e.g. \citealt{Kumar2021}), where the typical spectral width is $\frac{\delta \nu}{\nu} \approx 0.1$ \citep{Hessels2019}. Interestingly, the narrow spectral width of FRBs is a property also shared by galactic pulsar nanoshots \citep{Soglasnov2004,Hankins2007}, and other pulses from the Crab \citep{Thulasiram2021}.
\par
By inspection of Eq. \ref{eq:gammanu_c_curv}, we find that the frequency of curvature emission is determined by the lorentz factor of the radiating particles $\gamma$ and the curvature radius of the field lines $\rho$. In \cite{Wang2019}, the authors attribute the downward drifting of FRB sub-pulses, also known as the "sad trombone effect", to emission from particles propagating on field lines with different curvature radii. In this model, each FRB sub-burst originates from a chain of bunches on a single set of field lines. Chains of bunches propagating along different sets of field lines explain the spectral-temporal properties of FRBs with sub-bursts. Bunches that propagate on field lines with larger curvature radii (and thus radiate with a lower critical frequency) must travel further before becoming visible to the observer and thus are observed later in time, explaining the "sad trombone effect". 
\par
The authors also suggest the spectral width of each sub-burst could be described by the spread of each bunch's coherent particles' lorentz factors and/or the span of the curvature radii of field lines occupied by the bunch. Specifically, if we assume a constant lorentz factor $\gamma$ for all coherent emitting particles in a single bunch, then we can calculate the spectral width of an FRB using the geometry of the bunch. This means that $\frac{\delta \nu}{\nu} \approx \frac{\delta \rho}{\rho}$, where $\delta \rho$ is the difference in field line curvature radius as followed by electrons of the same bunch. This assumption requires that for each bunch $\big(\frac{\delta \gamma}{\gamma}\big)^3 \ll \frac{\delta \rho}{\rho}$, such that $\frac{\delta \rho}{\rho}$ is the dominant factor in determining the spectral width of radiation from a single bunch. This is plausible for two reasons. Firstly, differences in lorentz factors would lead to bunch dispersal along the direction of velocity and lead to the break-up of the bunch. Secondly, the lorentz factor of particles can be determined by balancing acceleration coherent losses, assuming continuous acceleration. Each particle emits coherently with a number of other surrounding particles based on spatial vicinity and similarity of velocity vectors. Particles that have a very high coherence factor (i.e. are emitting coherently with the largest number of other particles) contribute to the brightest emission and therefore the observed spectral width. These particles also undergo similar radiative losses and therefore share a common lorentz factor. 

\par
Let us assume a dipole magnetic field such that we can describe the curvature radius of field lines as a function of radial and polodial coordinates $(r,\theta)$ around the neutron star:
\begin{equation}
    \rho = \frac{r (1 + \cos^2(\theta))^{3/2}}{3 \sin(\theta) (1 + \cos^2(\theta))} \approx \frac{4 r}{3 \sin(\theta)}
    \label{eq:rho_theta}
\end{equation}
We assume a neutron star radius of $10^6$cm, and that coherent emission originates sufficiently close to the surface such that $r \approx 10^6$cm. The size of the bunches $R_t$ determines across which field lines coherent emission may span, i.e. $\delta \theta = R_t/r$ and $\delta r = \frac{R_l}{\gamma r} \approx \frac{c}{\nu r}$ in the NS frame. We can infer components of $\delta \nu$ due to longitudinal and transverse extents of the bunch. The transverse component is:
\begin{equation}
    \begin{split}
    \frac{\delta \rho_t}{\rho} &\approx \frac{\delta \theta}{\theta} = \frac{3 R_t \rho}{4 r^2} \\
    &\approx 0.1 \; R_{t,3} \, r_{6}^{-2} 
    \end{split}
    \label{eq:dnuovernu}
\end{equation}
Where we have assumed that $\theta \approx \frac{4 r}{3 \rho}$. 
The longitudinal component in the frame of the NS is:
\begin{equation}
    \begin{split}
    \frac{\delta \rho_l}{\rho} &\approx \frac{R_l}{r} \\
    &= \frac{c}{\nu r} \approx 3 \times 10^{-5} \; r_6^{-1} \\
    \end{split}
\end{equation}
Where we have assumed $R_l = \frac{c}{\nu_{\rm obs, curv}}$ in the frame of the neutron star. Therefore this component has a negligible contribution to $\frac{\delta \nu}{\nu}$, and the transverse component dominates. We are assuming here that the main component of the observed spectral width for each sub-burst comes from $\delta \nu$ of a single bunch. As each sub-burst must constitute a train of bunches each observable in succession, geometric differences between bunches could also contribute to the spectral width. Therefore we are implicitly assuming the chain of bunches along a set of field lines, which constitute one FRB sub-burst, have similar geometric properties. Furthermore, we have assumed that the radiation region is very close to the neutron star (i.e. $r = 10^6$cm). If the origin of bursts is a larger than few neutron star radii from the surface, as could be expected \citep{KumarBosnjak2020}, the span of the field lines $\delta \rho$ due to the width of the bunches $R_t$ would be too small to account for the observed spectral width. 
\par
\begin{figure}
    \centering
    \includegraphics[width=0.48\textwidth]{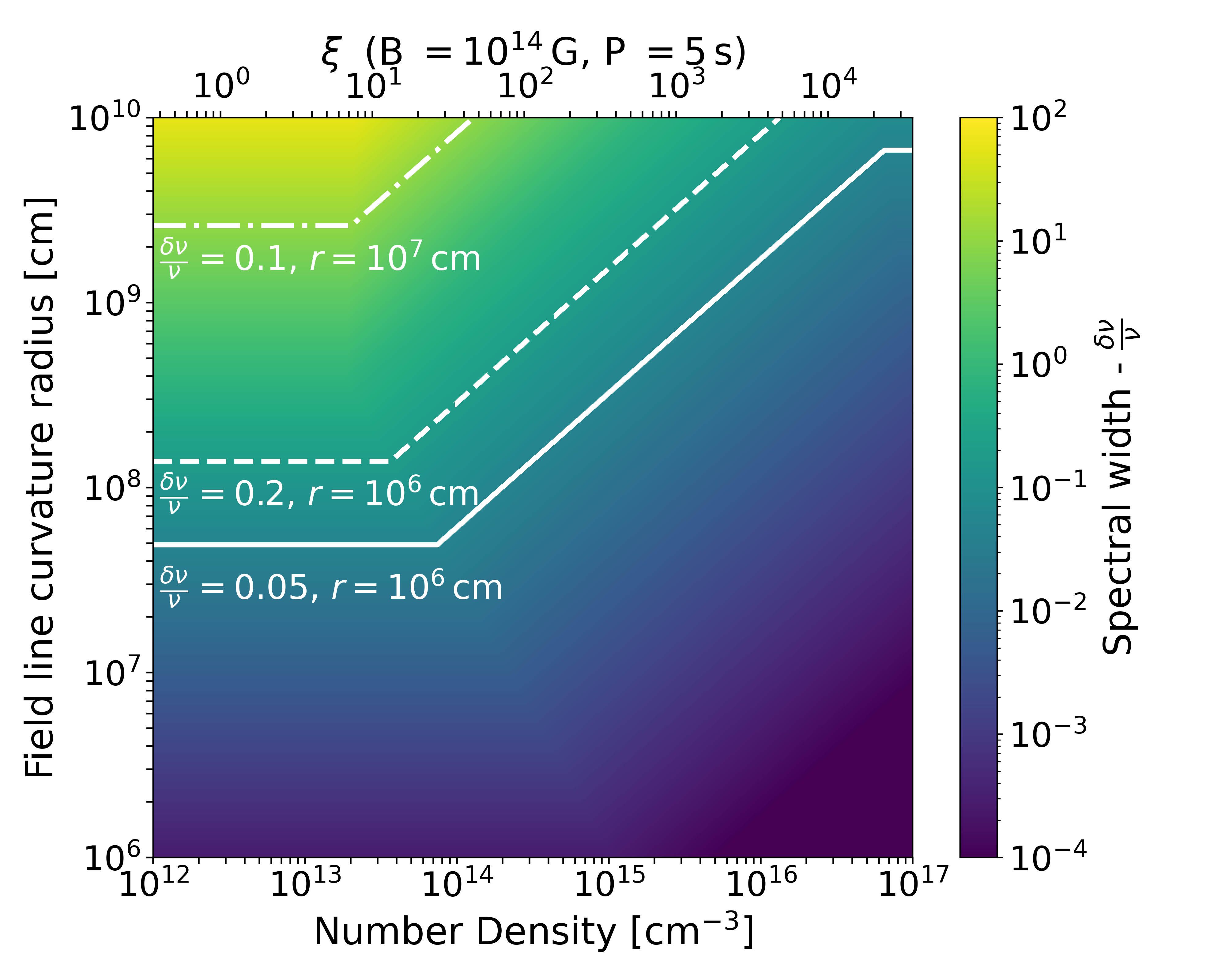}
    \caption{We show $\frac{\delta \nu}{\nu}$ from Eq. \ref{eq:dnuovernu} for a range of values of $\rho$ \& $n_e$. We have assumed $B = 10^{14} \, {\rm G}$, $\nu_{\rm obs, curv} = 10^9 \, {\rm Hz}$ and $l = {\rm min((R_t \rho)^{1/2}, 10^6 {\rm cm})}$. For emission originating on the surface of the magnetar, $r = 10^{6} \, {\rm cm}$, the solid white line refers to the contour defined by $\frac{\delta \nu}{\nu} = 0.05$, and the dashed line to $\frac{\delta \nu}{\nu} = 0.2$. We also plot the contour for $\frac{\delta \nu}{\nu} = 0.1$ assuming the emission originates at a distance $r = 10^{7} \, {\rm cm}$ from the surface. Only moderate number densities and large curvature radii allow a transverse size large enough such that observed values of $\frac{\delta \nu}{\nu}$ are reproduced. We note again that additional components to the spectral width may be present, e.g. if consecutively observable bunches have different geometric properties.}
    \label{fig:2ddeltanu}
\end{figure}

In Fig. \ref{fig:2ddeltanu} we plot the spectral width of emission due to Eq. \ref{eq:dnuovernu} in the $n_e$-$\rho$ parameter space. Due to the constraints upon $R_t$ discussed in this paper (Fig. 
\ref{fig:2d_bunch_size}), the observed spectral width of FRBs can only be reproduced on field lines with large curvature radii ($\rho \gtrsim 10^{8} \, {\rm cm}$). 
In Figs. \ref{fig:2ddeltanu}, the parameter space that corresponds to $0.05 <\frac{\delta \nu}{\nu} < 0.2$ via Eq. \ref{eq:dnuovernu} are over-plotted in white dashed and solid lines, assuming emission from the surface. This provides an insight into the allowed values of $\rho$ and $n_e$, e.g. $\rho \gtrsim 10^8 \,{\rm cm}$ and $n_e < 10^{15} \, {\rm cm^{-3}}$ are the preferred values in order to reproduce the observed FRB spectral width. The last open field line is defined as \citep{Sturrok1971}:
\begin{equation}
    \theta_p = \frac{r}{R_L} \approx 0.01 P^{-1/2}
\end{equation}
Where $R_L$ is the radius of the light cylinder. Solving for P, we find that a neutron star with a period of approximately $20$ seconds has a curvature radius of $10^8$cm at the polar cap. Given this and our results in Fig. \ref{fig:2d_spec_lum} we might expect longer period magnetars to be the source of FRBs with limited bandwidth, if emission originates from the polar cap.

\subsection{Bunch Luminosity}
\begin{figure}
    \centering
    \includegraphics[width=0.48\textwidth]{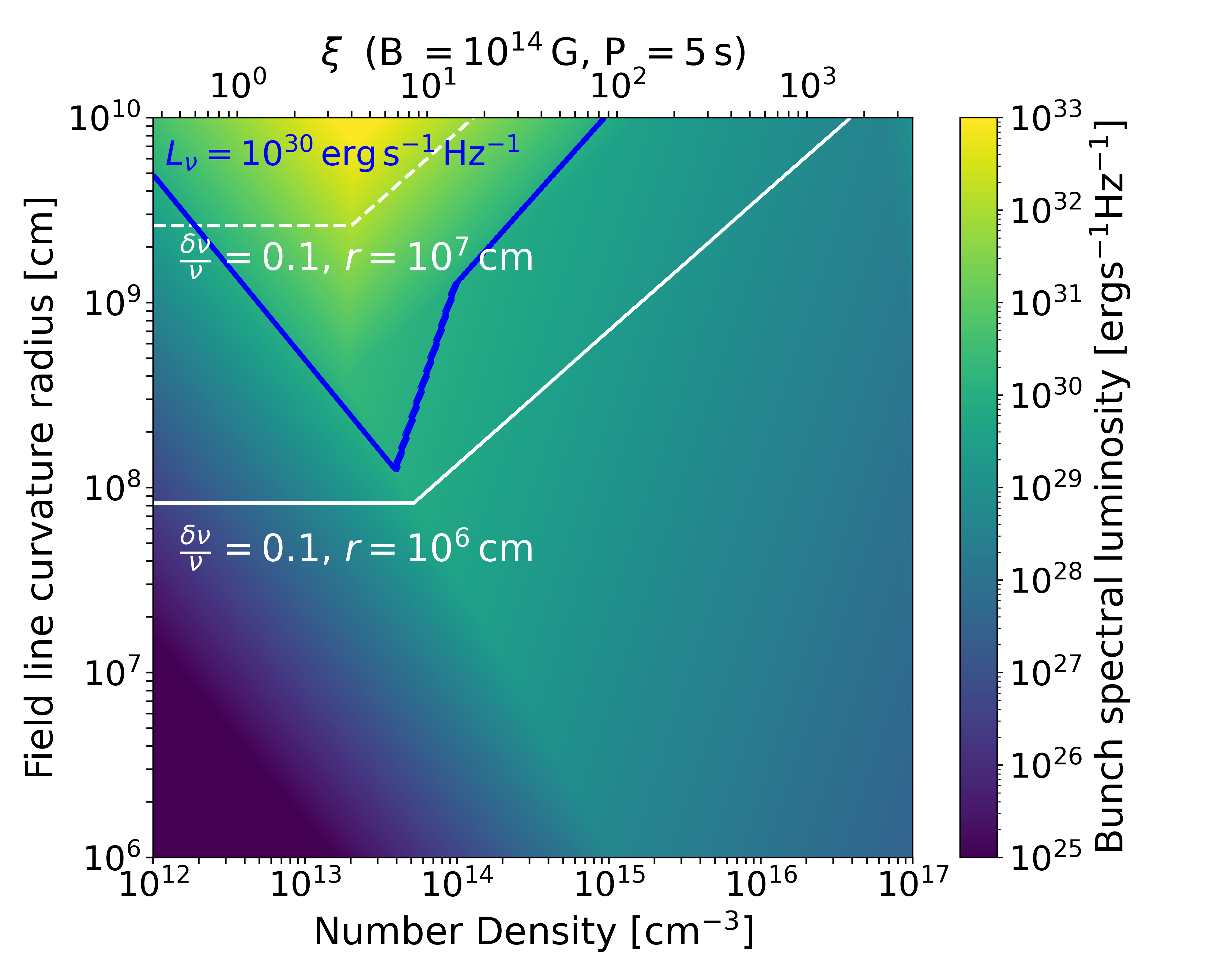}
    \caption{The maximum curvature luminosity of a bunch as a function of bunch density $n_e$ and field line curvature radius $\rho$. We have assumed $B = 10^{14} \, {\rm G}$, $\nu_{\rm obs, curv} = 10^9 \, {\rm Hz}$ and $l = {\rm min((R_t \rho)^{1/2}, 10^6 {\rm cm})}$. We again over-plot in white contours that refer to $\frac{\delta \nu}{\nu} = 0.1$ for bursts originating on the surface of the neuron star (solid), and from a distance of $r = 10^7$cm (dashed) from Sect. \ref{sect:spectral}. In blue we show the contour referring to a spectral luminosity of $L_{\nu} = 10^{30} \; {\rm erg \, s^{-1} \, Hz^{-1}}$. The parameter region from which FRB strength bursts can be produced is relatively limited, especially if spectral constraints are taken into account.}
    \label{fig:2d_spec_lum}
\end{figure}

The emitted luminosity per bunch assuming perfect coherence is $L_{\rm em} = N^2 P_{\rm curv, em} \propto \gamma^4$. The observed luminosity is $L_{\rm obs} = \gamma^2 N^2 P_{\rm curv, em} \propto \gamma^6$ due to relativistic beaming, and the isotropic equivalent luminosity is: $L_{\rm iso} = \gamma^4 N^2 P_{\rm curv, em} \propto \gamma^8$. We can assume from typical FRB spectra that the spectral luminosity of an FRB-emitting bunch is approximately $L_{\nu} \approx \frac{L}{0.1 \nu}$, and therefore we can write the observed spectral luminosity from one bunch as:
\begin{equation}
    \begin{split}
        L_{\nu, \rm obs} &= N^2 P_{\rm curv, obs} (0.1 \nu_{\rm obs, curv})^{-1}  \\
        &\propto \; n_e^2 \, R_t^4 \, \rho^{2/3} \, \nu_{\rm obs, curv}^{-1/3}
        \label{eq:spectral_lum}
    \end{split}
\end{equation}
Where once again we have eliminated $\gamma$ using Eq. \ref{eq:gammanu_c_curv}, and we have used the fact that the maximum longitudinal size of the bunch in its own frame is given by $R_l = \frac{\gamma c}{\nu_{\rm obs, curv}}$, i.e. the wavelength of radiation in the bunch's frame. We plot the maximum spectral luminosity due to Eq. \ref{eq:spectral_lum} in Fig. \ref{fig:2d_spec_lum}. Note that we assume here that the train of bunches are consecutively observable, however it is possible that coherent emission from multiple observable bunches add incoherently to a total luminosity. 
FRBs have a typical spectral luminosities of $10^{28-33} \; {\rm erg \, s^{-1} \, Hz^{-1}}$ and therefore this plot implies the $n_e$-$\rho$ parameter space from which FRBs can be produced is limited. A bunch number density of $10^{13-14} \; {\rm cm^{-3}}$ is preferred to produce FRB-strength bursts, corresponding to $\xi = 10-100$ for $B = 10^{14} \; {\rm G}$ and $P = 10 \; {\rm s}$. Cross-referencing with the spectral constraints imply a number density of around $10^{14} \; {\rm cm^{-3}}$ and a curvature radius $\rho \approx 10^8\; {\rm cm}$ for FRB-strength emission.  
A requirement of the coherent curvature radiation model of FRBs is that the local magnetic field should be $B \gtrsim 10^{14} \, {\rm G}$ \citep{Kumar2017}. We also require that the density of bunches should be greater than the local Goldreich-Julian density \citep{YangZhang2018}. Therefore by Eq. \ref{eq:goldreich}, we can rule out FRB-strength bursts from magnetars with short spin periods ($P < 0.01$s). The brightest FRBs require a magnetar spin period of at least $P > 1$s (corresponding to $\xi \approx 1$). This value is higher if either: the radiation region is located further away from the neutron star surface, or if $\xi > 1$. If the multiplicity of bunches powering FRBs is $\xi = 10^3$ as has been suggested in the literature, then the inferred minimum magnetar period for bright FRBs is $10^{2-3}$ seconds. Given this, we suggest that slowly rotating magnetars may be candidates of FRB sources.


\section{Discussion}
\label{sect:discussion}
\subsection{Extremely short duration emission}
In this paper we have discussed the quenching of coherent radio emission due to photon-pair cascades screening the electric field required for continuous acceleration which balance intense radiative losses. As discussed in Section \ref{sect:2}, the characteristic timescale upon which the cascades occur is $\tau_{\rm pp}$, or the one-photon pair creation timescale. This is approximately $l/c$, assuming other conditions are met, which is less than around $10^{-5} \,$s. Observations of the Crab Pulsar have shown that emission can range from millisecond to sub-nanosecond \citep{Hankins2007}. The Crab Pulsar is the most prolific example of such short bursts, and most extreme Crab nanoshot was observed with a peak flux of 2 Mega-Jansky, or approximately $L_{\nu} \approx 10^{28} \; {\rm erg \, s^{-1} \, Hz^{-1}}$; \citealt{Hankins2007}), with a maximum (unresolved) duration of $0.4$ nanoseconds. More recently, extra-Galactic FRBs have been observed with sub-millisecond durations, and isolated shots with structure on nanosecond timescales \citep{Nimmo2021}.
\par
\begin{figure}
    \centering
    \includegraphics[width=0.48\textwidth]{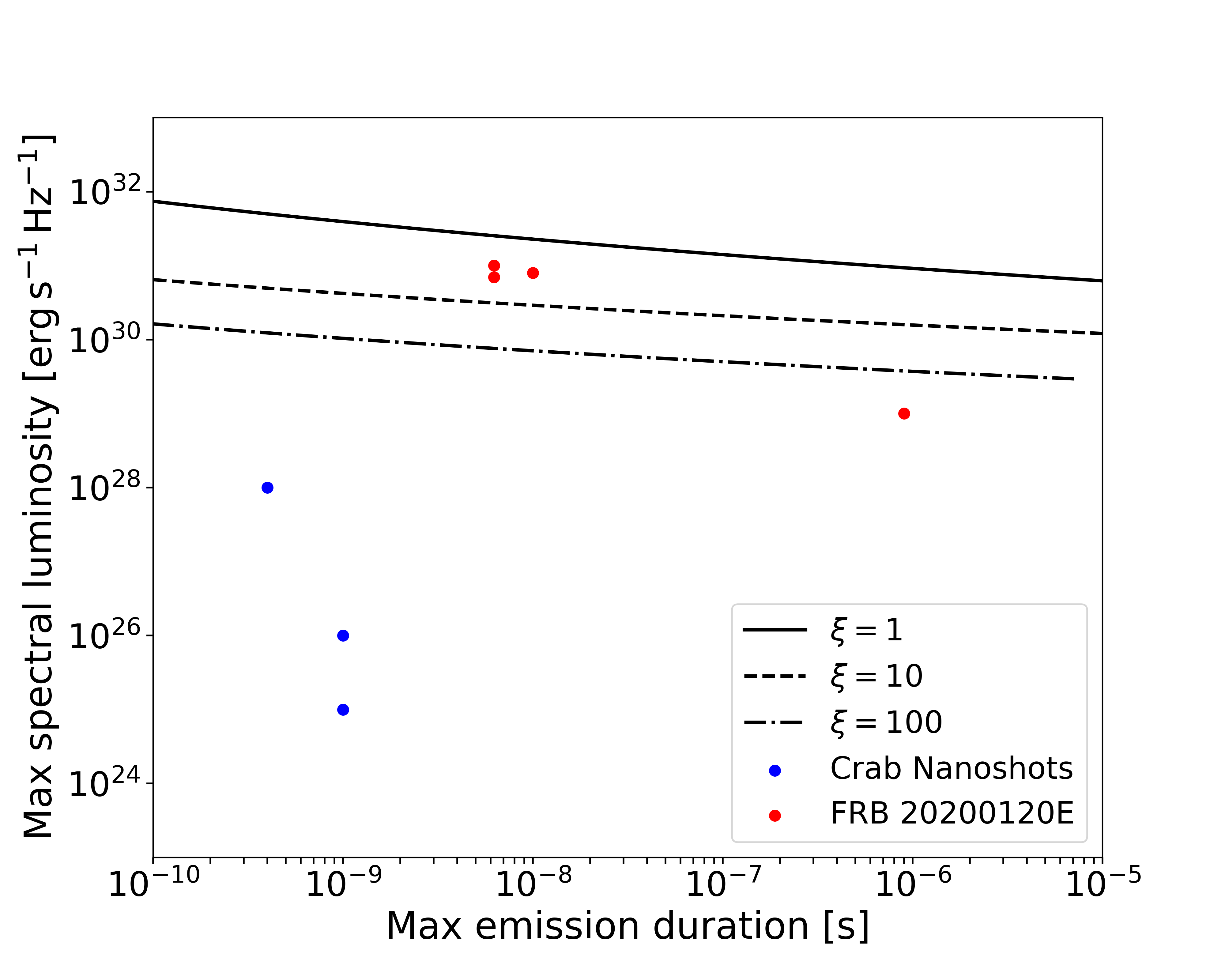}
    \caption{The maximum duration of bursts of different luminosities, for bursts from the surface of a magnetar with parameters: $P = 5 \,$s, $B =10^{14} $G; and $\rho = 10^8 \,$cm. We also plot the spectral luminosity and duration of Crab nanoshots \citep{Hankins2003,Hankins2007} and isolated shots from FRB20200120E \citep{Nimmo2021}.}
    \label{fig:luminosity_duration_crab}
\end{figure}

Although the short duration of these bursts may be intrinsic to the trigger event, an interesting prediction of this work is that short coherent bursts could be limited in duration due to quenching by pair-photon cascades. If we assume this to be the case, one can find a relation between the luminosity and duration of bursts due to cascade initiated quenching for bursts with durations less than $\frac{l}{c}$. Again we require that $T_{\rm pp} > l^{-1}$ where $l$ is defined in Eq. \ref{eq:l_definition}, such that pairs are produced in the same region as the bunches are accelerated. If this condition is met, the maximum burst duration is estimated as $\delta t = (c T_{\rm pp})^{-1}$.  In Fig. \ref{fig:luminosity_duration_crab} we plot the spectral luminosity of coherent curvature radiation bunches (using Eq. \ref{eq:spectral_lum}) from a typical magnetar, as a function of the maximum duration of a burst as limited by cascade-induced quenching. For reference, we also plot the timescale and luminosity of observed nanoshots and isolated shots from FRBs, which fall within the allowed luminosity-duration space for the stated parameters. 

\subsection{One-photon pair production that does not screen the electric field}
Bunches radiating coherent curvature emission may emit photons above the pair production threshold even if emission is not quenched. This is the case for example if $T_{\rm pp} < l^{-1}$, but the inequality in Eq. \ref{eq:photons_above_threshold} is still met. In this case, the photo-magnetic interaction site occurs on magnetic field lines where particle acceleration does not take place, and the effect is to load pairs onto surrounding field lines. The total number of secondary pairs produced, $N_{\rm sec}$, is related to the number of primary coherently emitting particles $N_{\rm pri}$ as: $N_{\rm sec} \approx 2 N_{\rm pri} \frac{t}{\tau_{\rm ph}} \approx 10^9 \, N_{\rm pri}$. Where we have used Eq. \ref{eq:tph} and have assumed a radiation time $t = 10^{-3}$s. We can estimate the value of $N_{\rm pri}$ given the results of this work. For FRB emission we require at least $10^6$ bunches (Sect. \ref{sect:cascades}) with a density of approximately $10^{13}\, {\rm cm^{-3}}$ (Fig. \ref{fig:2d_spec_lum}), each with a volume of approximately $V = R_t^2 R_l = R_t^2 c \nu_{\rm obs}^{-1} \approx 10^7 \, R_{t,3}^2 \, \nu_{\rm obs, 9}^{-1} \: {\rm cm^3}$. This gives us a total number of primaries of approximately $10^{29}$ in agreement with \cite{Cordes+2016}. The secondary pairs are produced at least within an length scale $\rho$ such that photons interact with the field, and therefore we can put a lower limit on the secondary pair density: $n_{\rm sec} > N_{\rm sec}/\rho^3 \approx 10^{14} \, \rho_8^{-3} \, {\rm cm^{-3}}$. The enhanced particle density due to secondary particles loading surrounding field lines may lead to observable emission at other wavelengths during or after the coherent burst.

\subsection{Lower magnetic fields}
We have primarily presented results for the maximum transverse size of bunches in magnetic fields where $B \approx 10^{14} \,$G, as these are typically required for FRB strength bursts from coherent curvature radiation \citep{Kumar2017}. The constraints discussed in this work on the size of coherently emitting bunches also apply to bunches in lower strength magnetic fields, provided high-energy photons have the required energy to produce pairs and initiate cascades. However, lower magnetic fields lead to less stringent constraints on $R_t$ and the bunch luminosity, as $\chi \propto \epsilon B$.
\begin{figure}
    \centering
    \includegraphics[width=0.48\textwidth]{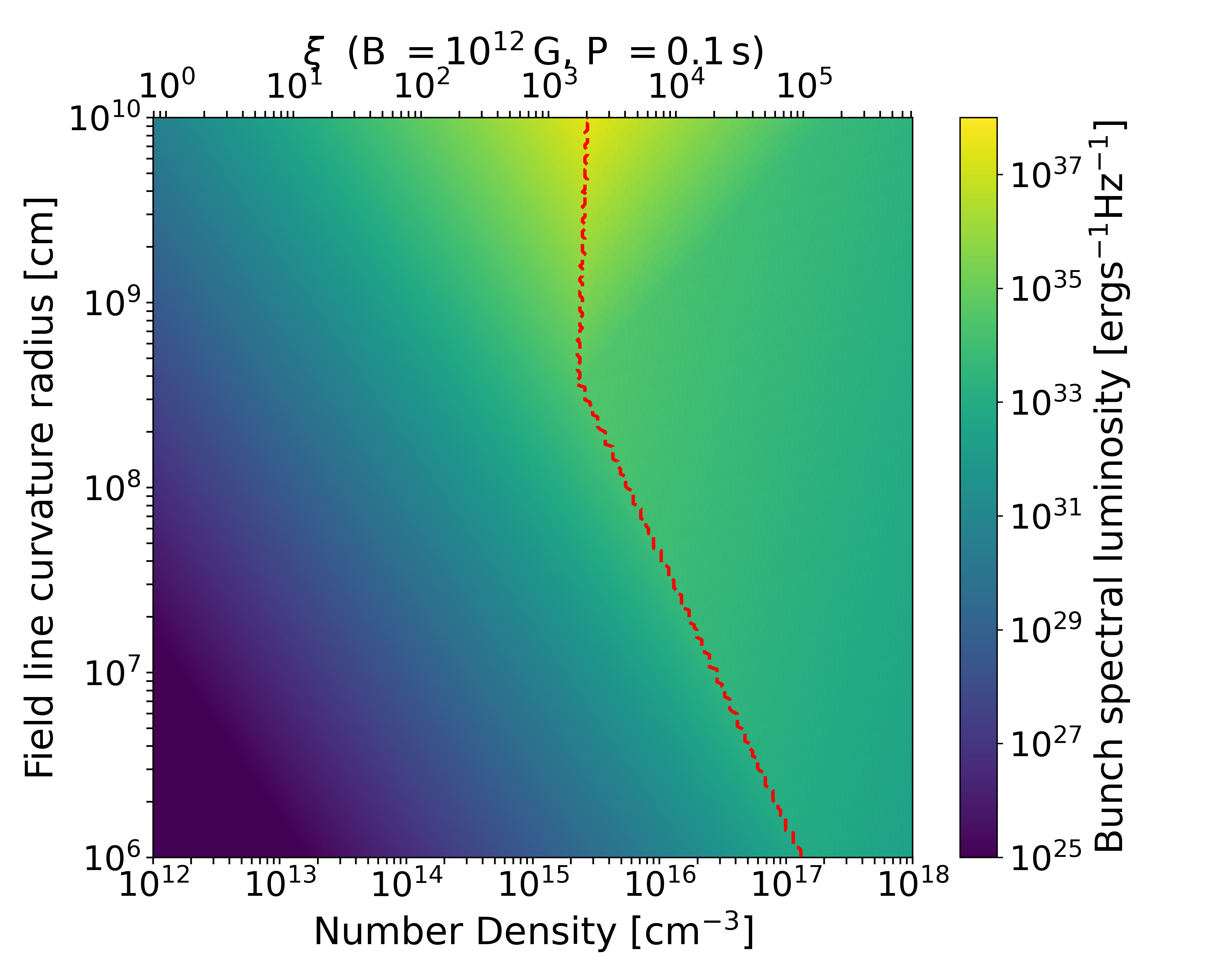}
    \caption{The maximum curvature luminosity of a bunch as a function of bunch density $n_e$ and field line curvature radius $\rho$ for a lower magnetic field of $B = 10^{12} \, {\rm G}$. For a lower magnetic field, photons must have higher energies to interact with the magnetic field to produce pairs and therefore the bunch size is less constrained. The cascade constraint is only limiting above $n_e = 2 \times 10^{15} \,{\rm cm^{-3}}$.}
    \label{fig:2d_spec_lum_lower_B}
\end{figure}
In Fig. \ref{fig:2d_spec_lum_lower_B} we plot the maximum bunch luminosity for $B = 10^{12} \,$G. 

\subsection{Other model predictions}
The growth of bunches by the radiation reaction requires a small spread in the angular velocities of bunch members \citep{GoldreichKeely1971}. The constraints discussed outlined in this work on the limited transverse size of FRB-emitting bunches may also limit differential angular velocities. Furthermore, our results suggest that the transverse size of bunches approximately depend inversely on the bunch density in the high density limit. Therefore in this regime, only a thin current sheet directly along the magnetic field lines is required for efficient bunch acceleration and radiation. The thickness of a current sheet may depend on the local plasma frequency as $l_{\rm cs} = \frac{\eta_{\rm cs} c}{\omega_{\rm p}} \approx 5 \times 10^{2} \; \eta_{\rm cs,3} \, n_{e,12}^{-1/2} \: {\rm cm}$, where $\eta_{\rm cs}$ values up to $10^3$ are suggested by simulations \citep{Sironi2016}. 

\section{Conclusion}
In this work, we have shown that the high-energy radiation that emerges from particle gyration in coherent bunches in the magnetosphere, as discussed in \cite{Cooper2021}, can affect the properties of the coherent curvature radio emission. In Section \ref{sect:2} we show that photo-magnetic processes lead to pair cascades which short the accelerating electric field and quench emission. In Section \ref{sect:consequences} we discuss the consequences of the pair cascades, and show that the nature of coherent bunches can be constrained due to these considerations. We discuss the maximum bunch size, and the affect this size has on the (spectral) luminosity of observed emission and the spectral properties of the emission. These results imply that in the coherent curvature radiation model, charge bunches powering FRBs propagate along magnetic field lines with a curvature radius $\rho \gtrsim 10^{8} \, {\rm cm}$, and have bunch densities $n_e \approx 10^{13-14} \, {\rm cm^{-3}}$. Assuming bunch densities must be above the local Goldreich-Julian density, our results suggest FRB-emitting magnetars have spin periods greater than 1 second, and possibly much higher. 

\section*{Acknowledgements}
AC is supported by the Netherlands Research School for Astronomy (NOVA). We thank D. Aksulu for comments on the draft of this manuscript. 

\section*{Data Availability}
A Python notebook from which the results and figures of this work can be reproduced will be made available upon publication. 


\bibliographystyle{mnras}
\bibliography{references} 





\bsp	
\label{lastpage}
\end{document}